\documentclass[12pt]{article}
\usepackage{amsmath,amssymb}
\newcommand{\eqdef}{\stackrel{\rm def}{=}}
\newcommand{\n}{\nonumber \\}

\begin{document}

\baselineskip=20pt

\newfont{\elevenmib}{cmmib10 scaled\magstep1}
\newcommand{\preprint}{
     \begin{flushleft}
       \elevenmib Yukawa\, Institute\, Kyoto\\
     \end{flushleft}\vspace{-1.3cm}
     \begin{flushright}\normalsize  \sf
       DPSU-06-2\\
       YITP-06-24
     \end{flushright}}
\newcommand{\Title}[1]{{\baselineskip=26pt
     \begin{center} \Large \bf #1 \\ \ \\ \end{center}}}
\newcommand{\Author}{\begin{center}
     \large \bf Satoru Odake${}^a$ and Ryu Sasaki${}^b$ \end{center}}
\newcommand{\Address}{\begin{center}
       $^a$ Department of Physics, Shinshu University,\\
       Matsumoto 390-8621, Japan\\
       ${}^b$ Yukawa Institute for Theoretical Physics,\\
       Kyoto University, Kyoto 606-8502, Japan
     \end{center}}

\preprint
\bigskip\bigskip\bigskip

\Title{Exact solution in the Heisenberg picture and 
annihilation-creation operators}

\Author

\Address

\begin{abstract}
The annihilation-creation operators of the harmonic oscillator,
the basic and most important tools in quantum physics,
are generalised to most solvable quantum mechanical systems of single
degree of freedom including the so-called `discrete' quantum mechanics. 
They admit exact Heisenberg operator solution.
We present unified definition of the annihilation-creation
operators ($a^{(\pm)}$) as the positive/negative frequency parts of
the exact Heisenberg operator solution.
\end{abstract}






\section*{Introduction}
\label{intro}
\setcounter{equation}{0}

The annihilation-creation operators are the simplest solution method
for quantum mechanical systems. In this Letter we provide
unified definition of annihilation-creation operators for most solvable
quantum mechanical systems of one degree of freedom.
A quantum mechanical system is called {\em solved\/} or {\em solvable\/}
if the entire
energy spectrum $\{\mathcal{E}_n\}$ and the corresponding eigenvectors
$\{\phi_n\}$, $\mathcal{H}\phi_n=\mathcal{E}_n\phi_n$ are known
\cite{susyqm}. This is the solution in the Schr\"odinger picture.
We will show that they also possess exact Heisenberg operator solutions.
The annihilation-creation operators are defined as the positive/negative
frequency parts of the Heisenberg operator solution 
and they are hermitian conjugate to each other.
This method also applies to the `discrete' quantum mechanical systems,
which are deformations of quantum mechanics obeying certain difference
equations; see full paper \cite{os7}.
Our results will be translated to those of the corresponding orthogonal
polynomials.
In particular, the exact Heisenberg operator solution of the
`sinusoidal coordinate'
corresponds to the so-called {\em structure relation\/} \cite{koorn}
for the orthogonal polynomials, 
including the Askey-Wilson and Meixner-Pollaczek polynomials \cite{koeswart}.

\subsection*{Heisenberg Operator Solution}
\label{qmechanics}

We will focus on the discrete energy levels, finite or infinite in number,
which are non-degenerate in one dimension.
For the majority of the solvable quantum systems \cite{susyqm},
the $n$-th eigenfunction has the following general structure
$\phi_n(x)=\phi_0(x) P_n(\eta(x))$, 
in which $\phi_0(x)$ is the {\em ground state\/} wavefunction
and $P_n(\eta(x))$ is an orthogonal polynomial of degree $n$ in
a real variable $\eta$.

Our main claim is that this $\eta(x)$ undergoes a
`{\em sinusoidal motion\/}' under the given Hamiltonian $\mathcal{H}$,
at the classical as well as quantum level.
The latter is simply the exact Heisenberg operator solution.
To be more specific, at the classical level we have
\begin{equation}
  \{\mathcal{H},\{\mathcal{H},\eta\}\}=-\eta\,
  R_0(\mathcal{H})-R_{-1}(\mathcal{H}).
  \label{twopoi}
\end{equation}
The two coefficients $R_0$ and $R_{-1}$ are, in general,
polynomials in the Hamiltonian $\mathcal{H}$. This leads to a simple
sinusoidal time-evolution:
\begin{align}
  & \ \ \eta(x;t)=\sum_{n=0}^\infty\bigl({(-t)^n/n!}\bigr)
  ({\rm ad}\,\mathcal{H})^n\eta\n 
  &=-\{\mathcal{H},\eta\}_0\,
  \frac{\sin\bigl[t\sqrt{R_0(\mathcal{H}_0)}\,\bigr]}
  {\sqrt{R_0(\mathcal{H}_0)}}
  -R_{-1}(\mathcal{H}_0)/R_{0}(\mathcal{H}_0)\n
  &\quad+\bigl(\eta(x)_0+R_{-1}(\mathcal{H}_0)/R_{0}(\mathcal{H}_0)\bigr)
  \cos\bigl[t\sqrt{R_0(\mathcal{H}_0)}\,\bigr],
  \label{clasol}
\end{align}
in which ${\rm ad}\,\mathcal{H}\,X=\{\mathcal{H},X\}$ and 
the subscript 0 means the initial value (at $t=0$).
The corresponding quantum expression is
\begin{equation}
  [\mathcal{H},[\mathcal{H},\eta]\,]=\eta\,R_0(\mathcal{H})
  +[\mathcal{H},\eta]\,R_1(\mathcal{H})+R_{-1}(\mathcal{H}),
  \label{twocom}
\end{equation}
in which $R_1(\mathcal{H})$ is the quantum effect.
The exact Heisenberg operator solution reads
\begin{align}
  & e^{it\mathcal{H}}\eta(x)e^{-it\mathcal{H}}
  =\sum_{n=0}^\infty\bigl({(it)^n/n!}\bigr)({\rm ad}\,\mathcal{H})^n\eta\n
  =\ &[\mathcal{H},\eta(x)]
  \frac{e^{i\alpha_+(\mathcal{H})t}-e^{i\alpha_-(\mathcal{H})t}}
  {\alpha_+(\mathcal{H})-\alpha_-(\mathcal{H})}
  -R_{-1}(\mathcal{H})/R_{0}(\mathcal{H})\n
  & +\bigl(\eta(x)+R_{-1}(\mathcal{H})/R_0(\mathcal{H})\bigr)
  \times\frac{-\alpha_-(\mathcal{H})e^{i\alpha_+(\mathcal{H})t}
  +\alpha_+(\mathcal{H})e^{i\alpha_-(\mathcal{H})t}}
  {\alpha_+(\mathcal{H})-\alpha_-(\mathcal{H})},
  \label{quantsol}
\end{align}
in which ad is now a commutator 
${\rm ad}\,\mathcal{H}\,X=[\mathcal{H},X]$, instead of the Poisson
bracket. The two ``frequencies'' are
\begin{gather}
  \alpha_\pm(\mathcal{H})=\bigl(R_1(\mathcal{H})\pm
  \sqrt{R_1(\mathcal{H})^2+4R_0(\mathcal{H})}\,\bigr)/2, \n
  \alpha_+(\mathcal{H})+\alpha_-(\mathcal{H})=R_1(\mathcal{H}), \n
  \alpha_+(\mathcal{H})\alpha_-(\mathcal{H})=-R_0(\mathcal{H}).
  \label{freqpm}
\end{gather}
If the quantum effects are neglected, {\em i.e.} $R_1\equiv0$ and
$\mathcal{H}\to\mathcal{H}_0$,
we have $\alpha_+=-\alpha_-=\sqrt{R_0(\mathcal{H}_0)}$, and the above
Heisenberg operator solution reduces to the classical one \eqref{clasol}.
When the exact operator solution \eqref{quantsol} is applied to $\phi_n$,
the r.h.s. has only three time-dependence, $e^{i\alpha_\pm(\mathcal{E}_n)t}$
and a constant. Thus the l.h.s. can only have two non-vanishing matrix
elements when sandwiched by $\phi_m$, except for the obvious $\phi_n$
corresponding to the constant term. In accordance with the above general
structure of the eigenfunctions, they are $\phi_{n\pm1}$;
that is $\langle\phi_m|\eta(x)|\phi_n\rangle=0$, for
$m\neq n\pm1,n$. This imposes the following conditions on the energy
eigenvalues
\[
  \mathcal{E}_{n+1}-\mathcal{E}_{n}=\alpha_+(\mathcal{E}_n),\quad
  \mathcal{E}_{n-1}-\mathcal{E}_{n}=\alpha_-(\mathcal{E}_n).
\]
These conditions, together with their `hermitian conjugate' ones, combined
with the ground state energy $\mathcal{E}_0=0$ determine the entire
discrete energy spectrum as shown by Heisenberg and Pauli.
In this letter we adopt the factorised Hamiltonian:
\[
  \mathcal{H}=\mathcal{A}^\dagger \mathcal{A}/2,\quad
  \mathcal{A}\phi_0=0 \Rightarrow
  \mathcal{H}\phi_0=0,\quad
  \mathcal{E}_0=0.
\]
The consistency of the procedure requires that the coefficient of
$e^{i\alpha_-(\mathcal{H})t}$ on the r.h.s. should vanish when applied
to the ground state $\phi_0$:
\[
  -[\mathcal{H},\eta(x)]\phi_0+\bigl(\eta(x)\alpha_+(0)
  -R_{-1}(0)/\alpha_-(0)\bigr)\phi_0=0,
\]
which is the equation determining the ground
state eigenvector $\phi_0$ in the Heisenberg picture.

Thus we arrive at a {\em dynamical and unified  definition\/} of
the {\bf annihilation-creation operators}:
\begin{align}
  e^{it\mathcal{H}}\eta(x)e^{-it\mathcal{H}}
  &=a^{(+)}(\mathcal{H},\eta)e^{i\alpha_+(\mathcal{H})t}
  -R_{-1}(\mathcal{H})/R_{0}(\mathcal{H})
  +a^{(-)}(\mathcal{H},\eta)e^{i\alpha_-(\mathcal{H})t},
  \label{acdefs0}
\end{align}
\vspace{-20pt}
\begin{align}
  a^{(\pm)}&=a^{(\pm)}(\mathcal{H},\eta)\n
  &\eqdef\Bigl(\pm[\mathcal{H},\eta(x)]
  \mp\!\bigl(\eta(x)+R_{-1}(\mathcal{H})/R_0(\mathcal{H})\bigr)
  \alpha_\mp(\mathcal{H})\Bigr)\!\bigm/\!
  \bigl(\alpha_+(\mathcal{H})-\alpha_-(\mathcal{H})\bigr).
  \label{acdefs}
\end{align}
When acting on the eigenvector $\phi_n$, they read
\begin{align}
  a^{(\pm)}\phi_n(x)&=\frac{\pm1}{\mathcal{E}_{n+1}-\mathcal{E}_{n-1}}
  \Bigl([\mathcal{H},\eta(x)]
  +(\mathcal{E}_n-\mathcal{E}_{n\mp 1})\eta(x)
  +\frac{R_{-1}(\mathcal{E}_n)}{\mathcal{E}_{n\pm 1}-\mathcal{E}_n}\Bigr)
  \phi_n(x).
\end{align}
By using the three-term recursion relation of the orthogonal polynomial
$P_n$
\[
  \eta P_n(\eta)=A_n P_{n+1}(\eta)+B_n P_n(\eta)+C_n P_{n-1}(\eta)
\]
on the l.h.s. of \eqref{acdefs0}, we arrive at
\begin{equation}
  a^{(+)}\phi_n=A_n\phi_{n+1},\quad a^{(-)}\phi_n=C_n\phi_{n-1}.
  \label{acphi=}
\end{equation}
Based on these relations, it is easy to show that $a^{(\pm)}$ are
{\em hermitian conjugate\/} to each other.
Sometimes it is convenient to introduce $a^{\prime(\pm)}$  with a different
normalisation
\begin{align}
  a^{\prime(\pm)}&\eqdef a^{(\pm)}
  \bigl(\alpha_+(\mathcal{H})-\alpha_-(\mathcal{H})\bigr)
  \label{ac'defs}\\
  &=\pm[\mathcal{H},\eta(x)]
  \mp\bigl(\eta(x)+R_{-1}(\mathcal{H})/R_0(\mathcal{H})\bigr)
  \alpha_\mp(\mathcal{H}),
  \nonumber
\end{align}
which are no longer hermitian conjugate to each other.

In the literature there is a quite wide variety of proposed annihilation
and creation operators \cite{coherents}. Historically most of these
operators are connected to the so-called
{\em algebraic theory of coherent states\/}, which are defined as
the eigenvectors of the annihilation operator (AOCS, Annihilation
Operator Coherent State).
Therefore, for a given potential or a quantum Hamiltonian,
there could be as many coherent states as the definitions of the
annihilation operators. In our theory, on the contrary, the
annihilation-creation operators are uniquely determined except for the
overall normalisation.
In terms of the simple parametrisation
$\psi=\sum_{n=0}^\infty c_n \phi_n(x)$ the equation
$a^{(-)}\psi=\lambda\psi$, $\lambda\in\mathbb{C}$ can be solved
with the help of the formula \eqref{acphi=}
\begin{equation}
  \psi=\psi(\lambda,x)=\phi_0(x)\sum_{n=0}^{\infty}
  \frac{\lambda^n}{\prod_{k=1}^nC_n}\cdot P_n(\eta(x)).
  \label{cohe}
\end{equation}

To the best of our knowledge, the `sinusoidal coordinate' was first
introduced in a rather broad sense for general (not necessarily
solvable) potentials as a useful means for coherent state research
by Nieto and Simmons \cite{nieto}.
In \cite{os7} the necessary and sufficient condition for the existence
of the `sinusoidal coordinate' \eqref{twocom} is analysed within the
context of ordinary quantum mechanics $\mathcal{H}=p^2/2+V(x)$.
It turns out that the potential can be expressed in terms of the
$\eta(x)$ and $d\eta/dx$
\begin{equation}
  V(x)=
  \Bigl({r^{(0)}_0}\eta^2/{2}+r^{(0)}_{-1}\eta
  +c\Bigr)/{({d\eta}/{dx})^2}-{r_1}/{8}.
  \label{cond3p}
\end{equation}
The parameters appear in $R_0(\mathcal{H})=r_0^{(1)}\mathcal{H}+r_0^{(0)}$,
$R_1(\mathcal{H})=r_1$ and
$R_{-1}(\mathcal{H})=r_{-1}^{(1)}\mathcal{H}+r_{-1}^{(0)}$
and $c$ is the constant of integration.
These potentials are all {\em shape invariant\/} \cite{genden}.
The Kepler problems in various coordinates and the Rosen-Morse 
potential are not
contained in \eqref{cond3p}, though they are shape invariant and solvable.

We give three typical examples. See \cite{os7} for more.

\subsubsection*{P\"oschl-Teller potential}
\label{poshtell}

The Hamiltonian of the P\"oschl-Teller potential, the eigenvalues and
the eigenfunctions are ($0<x<\pi/2$):
\begin{align}
  &\mathcal{H}\eqdef(p-ig\cot x+ih\tan x)(p+ig\cot x-ih\tan x)/2,\n
  &\mathcal{E}_n=2n(n+g+h),\quad g,h>0,
  \quad \eta(x)=\cos 2x,\n
  &\phi_n(x)=(\sin x)^g(\cos x)^hP_n^{(\alpha,\beta)}(\cos 2x),
\end{align}
in which $P_n^{(\alpha,\beta)}(\eta)$ is the Jacobi polynomial
and $\alpha=g-1/2$, $\beta= h-1/2$.
The classical solution of the initial value problem is
($\mathcal{H}'\eqdef\mathcal{H}+(g+h)^2/2$):
\begin{align}
  & \cos 2x(t)=\Bigl(\cos 2x(0)+\frac{g^2-h^2}{2\mathcal{H}'_0}\Bigr)
  \cos\bigl[2t\sqrt{2\mathcal{H}'_0}\,\bigr]\n
  &\quad\ -p(0)\sin 2x(0)
  \frac{\sin\bigl[2t\sqrt{2\mathcal{H}'_0}\,\bigr]}
  {\sqrt{2\mathcal{H}'_0}}-\frac{g^2-h^2}{2\mathcal{H}'_0}\,.
\end{align}
The corresponding quantum expressions are
\begin{align}
%
  \lbrack \mathcal{H},[\mathcal{H},\cos 2x]\,]&=
  \cos 2x\,(8\mathcal{H}'-4)
  +4[\mathcal{H},\cos 2x]
  +4(\alpha^2-\beta^2),
\end{align}
with $\alpha_\pm(\mathcal{H})=2\pm 2\sqrt{2\mathcal{H}'}$.
The annihilation and creation operators are
\begin{align}
  &a^{\prime(\pm)}/2=a^{(\pm)}2\sqrt{2\mathcal{H}'}
  =\pm \sin 2x\frac{d}{dx}+\cos 2x\,\sqrt{2\mathcal{H}'}
  +\frac{\alpha^2-\beta^2}{\sqrt{2\mathcal{H}'}\pm 1}\,.
\end{align}
When applied to the eigenvector $\phi_n$ as
$2\mathcal{E}_n+(g+h)^2=(2n+g+h)^2$, we obtain:
\begin{align}
  a^{\prime(-)}/2\,\phi_n
  &=\frac{4(n+\alpha)(n+\beta)}{2n+\alpha+\beta}\,\phi_{n-1},
  \label{PTdown}\\
  a^{\prime(+)}/2\,\phi_n
  &=\frac{4(n+1)(n+\alpha+\beta+1)}{2n+\alpha+\beta+2}\,\phi_{n+1}.
\end{align}

\subsubsection*{Deformed harmonic oscillator}
\label{defosci}

The deformed harmonic oscillator is a simplest example of shape invariant
`discrete' quantum mechanics.
The Hamiltonian of `discrete' quantum mechanics studied in this letter
has the following form \cite{os4} (with some modification for
the Askey-Wilson case):
\begin{align}
  &\mathcal{H}\eqdef\Bigl(\sqrt{V(x)}\,e^{\,p}\sqrt{V(x)^*}
  +\sqrt{V(x)^*}\,e^{-p}\sqrt{V(x)}
  -V(x)-V(x)^*\Bigr)\!\bigm/2.
  \label{H}
\end{align}
The eigenvalue problem for $\mathcal{H}$,
$\mathcal{H}\phi=\mathcal{E}\phi$ is a difference equation,
instead of a second order differential equation.
Let us define $S_{\pm}$, $T_{\pm}$ and $\mathcal{A}$ by
\begin{align}
  &S_+\eqdef e^{p/2}\sqrt{V(x)^*},\quad
  S_-\eqdef e^{-p/2}\sqrt{V(x)},\n
%
  &T_+\eqdef S_+^{\dagger}S_+=\sqrt{V(x)}\,e^p\sqrt{V(x)^*},\n
  &T_-\eqdef S_-^{\dagger}S_-=\sqrt{V(x)^*}\,e^{-p}\sqrt{V(x)},\n
%
  &\mathcal{A}\eqdef i(S_+-S_-),\quad
  \mathcal{A}^{\dagger}=-i(S_+^{\dagger}-S_-^{\dagger}).
  \label{aaddef}
\end{align}
Then the Hamiltonian is factorized
\begin{align}
  \mathcal{H}&=\bigl(T_++T_--V(x)-V(x)^*\bigr)/2\n
  &=(S_+^{\dagger}-S_-^{\dagger})(S_+-S_-)/2
  =\mathcal{A}^{\dagger}\mathcal{A}/2.
  \label{anotherfactham}
\end{align}
The potential function $V(x)$ of the deformed harmonic oscillator is
$V(x)=a+ix$, $-\infty<x<\infty$, $a>0$.
As shown in some detail in our previous paper \cite{os4},
it has an equi-spaced spectrum ($\mathcal{E}_n=n$, $n=0,1,2,\ldots$)
and the corresponding eigenfunctions are a special case of
the Meixner-Pollaczek polynomial $P_n^{(a)}(x\,;\tfrac{\pi}{2})$
\cite{koeswart},
\begin{align}
  \phi_0(x)&=\sqrt{\Gamma(a+ix)\Gamma(a-ix)},\quad \eta(x)=x,
  \label{phi0MP}\\
  \phi_n(x)&=\phi_0(x)P_n(x),\quad
  P_n(x)\eqdef P_n^{(a)}(x\,;\tfrac{\pi}{2}),
  \label{phinMP}
\end{align}
which could be considered as a deformation of the Hermite polynomial.

The Poisson bracket relations are
$\{\mathcal{H},x\}=-\sqrt{a^2+x^2}\,\sinh p$,
$\{\mathcal{H},\{\mathcal{H},x\}\}=-x$,
leading to the harmonic oscillation,
\begin{equation}
  x(t)=x(0)\cos t+\sqrt{a^2+x^2(0)}\,\sinh p(0)\, \sin t,
\end{equation}
which endorses the naming of the deformed harmonic oscillator.
The corresponding quantum expressions are also simple:
$[\mathcal{H},x]=-i(T_+-T_-)/2$, $[\mathcal{H},[\mathcal{H},x]\,]=x$,
\begin{align}
  e^{it\mathcal{H}}\,x\,e^{-it\mathcal{H}}&= x\,\cos t
  +i[\mathcal{H},x]\,\sin t
  = x\,\cos t+(T_+-T_-)/2\sin t.
\end{align}
The annihilation and creation operators are
\begin{eqnarray}
  a^{\prime(\pm)}=2a^{(\pm)}=x\pm[\mathcal{H},x]=x\mp i(T_+-T_-)/2.
\end{eqnarray}
These operators were also introduced in 
\cite{degruij} by a different reasoning from ours.
The action of the annihilation creation operators on the eigenvectors is
\begin{equation}
  a^{\prime(-)}\phi_n=(n+2a-1)\phi_{n-1},\
  a^{\prime(+)}\phi_n=(n+1)\phi_{n+1}.
  \label{mxpoldown}
\end{equation}
{}From these it is easy to verify the $\mathfrak{su}(1,1)$ commutation
relations including the Hamiltonian $\mathcal{H}$:
\begin{equation}
  [\mathcal{H},a^{\prime(\pm)}]=\pm\, a^{\prime(\pm)},\quad
  [a^{\prime(-)},a^{\prime(+)}]=2(\mathcal{H}+a).
\end{equation}

The coherent state \eqref{cohe}, is simply obtained
from the formula \eqref{mxpoldown} and $a^{\prime(-)}=2a^{(-)}$:
\begin{equation}
  \psi(x)=\phi_0(x)\sum_{n=0}^\infty\frac{(2\lambda)^n}{(2a)_n}
  P_{n}^{(a)}(x\,;\tfrac{\pi}{2}),
\end{equation}
which has a concise expression in terms of the hypergeometric function
${}_1F_1$
\begin{eqnarray}
  \psi(x)=\phi_0(x)\,e^{2i\lambda}\,{}_1F_1\Bigl(
  \genfrac{}{}{0pt}{}{a+ix}{2a}
  \Bigm|-4i\lambda\Bigr).
\end{eqnarray}

\subsubsection*{Askey-Wilson polynomial}
\label{aswil}

The Askey-Wilson polynomial belongs to the so-called $q$-scheme of
hypergeometric orthogonal polynomials \cite{koeswart}. It has four parameters
$a_1,a_2,a_3,a_4$ on top of $q$ ($0<q<1$), and is considered as a
three-parameter deformation of the Jacobi polynomial.
As a dynamical system, it could be called a deformed
P\"oschl-Teller potential.
The quantum-classical correspondence has some subtlety
because of another `classical' limit $q\to1$.

The Hamiltonian of the Askey-Wilson polynomial
has a bit different form from 
\eqref{H}:
\begin{align}
  \mathcal{H}\eqdef\Bigl(\sqrt{V(z)}\,q^{D}\!\sqrt{V(z)^*}
  +\sqrt{V(z)^*}\,q^{-D}\!\sqrt{V(z)}
  -V(z)-V(z)^*\Bigr)\!\bigm/2,
  \label{asH}
\end{align}
with a potential function $V(z)$, $z=e^{ix}$, $0<x<\pi$\,:
\[
   V(z)=\frac{\prod_{j=1}^4(1-a_jz)}{(1-z^2)(1-qz^2)}\,,\
   D\eqdef z\frac{d}{dz}=-i\frac{d}{dx}=p.
\]
We assume $-1<a_1,a_2,a_3,a_4<1$ and $a_1a_2a_3a_4<q$.
The eigenvalues and eigenfunctions are \cite{os4}:
\begin{align}
  \mathcal{E}_n&=(q^{-n}-1)(1-a_1a_2a_3a_4q^{n-1})/2,\n
  \phi_0(x)&=\sqrt{
  \frac{(z^2\,;q)_{\infty}}{\prod_{j=1}^4(a_jz\,;q)_{\infty}}
  \frac{(z^{-2}\,;q)_{\infty}}{\prod_{j=1}^4(a_jz^{-1}\,;q)_{\infty}}}\,,
  \n
  \eta(x)&=(z+z^{-1})/{2}=\cos x,\ \phi_n(x)=\phi_0(x)P_n(\cos x),\n
%
  \quad
  P_n(\eta)&\eqdef p_n(\eta\,;a_1,a_2,a_3,a_4|q),
  \label{phinAW}
\end{align}
in which $p_n(\eta\,;a_1,a_2,a_3,a_4|q)$ is the Askey-Wilson polynomial
\cite{koeswart}.

The classical sinusoidal motion \eqref{clasol} holds
with the Hamiltonian ($\gamma=\log q$):
\begin{align}
  \mathcal{H}_c&=\sqrt{V_c(z)V_c(z)^*}\cosh\gamma p
  -\bigl(V_c(z)+V_c(z)^*\bigr)/2,\n
  &\quad V_c(z)={\prod_{j=1}^4(1-a_jz)}/{(1-z^2)^2}.
\end{align}
The coefficients in the classical expression are
\begin{align}
  R_0(\mathcal{H}_c)&=\gamma^2(\mathcal{H}_c^2+c_1\mathcal{H}_c+c_2),
  \n
  R_{-1}(\mathcal{H}_c)&=-\gamma^2(c_3\mathcal{H}_c+c_4),
  \nonumber
\end{align}
with $c_1=1+b_4$, $c_2=(1-b_4)^2/4$, $c_3=(b_1+b_3)/4$,
$c_4=(1-b_4)(b_1-b_3)/8$.
Here we use the abbreviation
\[
   b_1\eqdef\sum_{1\leq j\leq 4}a_j\,,\
   b_3\eqdef\!\!\!\sum_{1\leq j<k<l\leq 4}a_ja_ka_l\,,\
   b_4\eqdef\prod_{j=1}^4a_j\,.
\]

The corresponding quantum expressions are
\begin{align}
  R_0(\mathcal{H})&=q(q^{-1}-1)^2\Bigl((\mathcal{H}')^2
  -(1+q^{-1})^2b_4/4\Bigr)\,,\n
  R_1(\mathcal{H})&=q(q^{-1}-1)^2\,\mathcal{H}', \quad
  \mathcal{H}'\eqdef\mathcal{H}+(1+q^{-1}b_4)/2,\n
  R_{-1}(\mathcal{H})&=-q(q^{-1}-1)^2\Bigl((b_1+q^{-1}b_3)\mathcal{H}/4
  +(1-q^{-2}b_4)(b_1-b_3)/8\Bigr).
\end{align}
The two frequencies are:
\begin{align}
  &\alpha_{\pm}(\mathcal{H})
  =(q^{-1}-1)\Bigl((1-q)\mathcal{H}'
  \pm(1+q)\sqrt{(\mathcal{H}')^2-q^{-1}b_4}\,\Bigr)\!\bigm/2.
\end{align}
The annihilation-creation operators are:
\begin{align}
  a^{(\pm)}&=\Bigl(
  \pm(q^{-1}-1)\bigl(z^{-1}(1-qz^2)T_++z(1-qz^{-2})T_-\bigr)/4\n
  &\qquad \mp\cos x\,\alpha_{\mp}(\mathcal{H})
  \pm R_{-1}(\mathcal{H})\alpha_{\pm}(\mathcal{H})^{-1}\Bigr)
  \bigm/\bigl(\alpha_+(\mathcal{H})-\alpha_-(\mathcal{H})\bigr)\,.
\end{align}
Their effects on the eigenvectors are:
\begin{align}
  \!a^{(-)}\phi_n&
  =\frac{(1-q^n)\prod_{1\leq j<k\leq 4}(1-a_ja_kq^{n-1})}
  {2(1-b_4q^{2n-2})(1-b_4q^{2n-1})}\,\phi_{n-1},\\
  \!a^{(+)}\phi_n&=\frac{1-b_4q^{n-1}}
  {2(1-b_4q^{2n-1})(1-b_4q^{2n})}\,\phi_{n+1}.
\end{align}
The coherent state is
\begin{equation}
  \psi(x)=\phi_0(x)\sum_{n=0}^{\infty}
  \frac{(2\lambda)^n}{(q\,;q)_n}\,
  \frac{(a_1a_2a_3a_4\,;q)_{2n}}{\prod_{1\leq j<k\leq 4}(a_ja_k\,;q)_n}\,
  P_n(\cos x)\,.
\end{equation}

\paragraph{\bf Conclusions}

We have shown that most solvable quantum mechanics of one degree of freedom
have exact Heisenberg operator solution.
The annihilation-creation operators ($a^{(\pm)}$) are defined
as the positive/negative frequency parts of
the exact Heisenberg operator solution. These ($a^{(\pm)}$) are
hermitian conjugate to each other. This method also applies to the 
so-called `discrete' quantum mechanics
whose eigenfunctions are deformations of the classical orthogonal
polynomials known as the Askey-scheme of hypergeometric orthogonal
polynomials.

We thank F.\, Calogero for stimulating discussion.
This work is supported in part by Grant-in-Aid for Scientific
Research from the Ministry of Education, Culture, Sports, Science and
Technology, No.18340061 and No.16340040.


\end{document}